\documentclass[english,aps, prb, preprint, superscriptaddress, floatfix]{revtex4}
\usepackage[T1]{fontenc}
\usepackage[latin1]{inputenc}
\usepackage{graphicx}
\begin{document}

\title{Electronic Structures of SiC Nanoribbons}

\author{Lian Sun}
\affiliation{Hefei National Laboratory for Physical Sciences at
Microscale, University of Science and Technology of China, Hefei,
Anhui 230026, P.R. China}

\author{Yafei Li}
\affiliation{Institute of New Energy Material Chemistry, Institute
of Scientific Computing, Nankai University, Tianjin 300071, P. R.
China}

\author{Zhenyu Li}
\affiliation{Hefei National Laboratory for Physical Sciences at
Microscale, University of Science and Technology of China, Hefei,
Anhui 230026, P.R. China}

\author{Qunxiang Li}
\affiliation{Hefei National Laboratory for Physical Sciences at
Microscale, University of Science and Technology of China, Hefei,
Anhui 230026, P.R. China}

\author{Zhen Zhou}
\affiliation{Institute of New Energy Material Chemistry, Institute
of Scientific Computing, Nankai University, Tianjin 300071, P. R.
China}

\author{Zhongfang Chen}
\affiliation{Department of Chemistry, Institute for Functional
Nanomaterials, University of Puerto Rico, San Juan, PR 00931-3346,
USA}

\author{Jinlong Yang}\thanks{Corresponding author. E-mail: jlyang@ustc.edu.cn}
\affiliation{Hefei National Laboratory for Physical Sciences at
Microscale, University of Science and Technology of China, Hefei,
Anhui 230026, P.R. China}

\author{J. G. Hou}
\affiliation{Hefei National Laboratory for Physical Sciences at
Microscale, University of Science and Technology of China, Hefei,
Anhui 230026, P.R. China}

\date{\today}

\begin{abstract}
Electronic structures of SiC nanoribbons have been studied by
spin-polarized density functional calculations. The armchair
nanoribbons are nonmagnetic semiconductor, while the zigzag
nanoribbons are magnetic metal. The spin polarization in zigzag SiC
nanoribbons is originated from the unpaired electrons localized on
the ribbon edges. Interestingly, the zigzag nanoribbons narrower
than $\sim$4 nm present half-metallic behavior. Without the aid of
external field or chemical modification, the metal-free
half-metallicity  predicted for narrow SiC zigzag nanoribbons opens
a facile way for nanomaterial spintronics applications.

\end{abstract}

\pacs{73.22.-f, 73.61.Wp}

\maketitle

\section{INTRODUCTION}

Stimulated by the successful preparation of graphene, an intense
research interest has been focused on two dimensional (2D) layered
structures. Among several group IV elements, only carbon can take
either $sp^2$ or $sp^3$ bond configurations and form 2D layered
structure. Silicon prefers $sp^3$ instead of $sp^2$ hybridization.
As a result, it is difficult to construct stable Si structures of
fullerenes, nanotubes, and graphene-like sheets. \cite{Khan}
However, it is possible to form layered structure by mixing C and
Si. Recently, SiC single-wall nanotubes (SWNTs) with different
chiralities, diameters, and atomic configurations have been studied
by first principles calculations, \cite{Miyamoto, Menon, Zhao, Gali,
Baumeier, Baierle1, Baierle2} and synthesized experimentally via the
reaction of silicon with multiwalled carbon nanotubes at different
temperatures. \cite{Sun} In contrast to carbon SWNTs, SiC SWNTs are
always semiconducting independent of the helicity, with a direct
band gap for zigzag tubes and an indirect gap for armchair and
chiral tubes. It is suggested that SiC SWNTs are candidates for
nanodevices that operate in high-power, high-frequency and
high-temperature regimes. \cite{Harris, Pan, Feng}

Nanoribbon (NR) is an important structure for 2D layered materials.
Graphene NRs have been extensively studied in the last two years.
\cite{Fujita, Yuanbozhang, Berger, Son1, Son2, NL07, APL-kaner,
Jiang07, JACS-kaner, ZuanyiLi-PRL} The hydrogen passivated graphene
NRs have a nonzero band gap. The zigzag graphene NRs have a magnetic
insulating ground state with ferromagnetic ordering at each zigzag
edge and antiparallel spin orientation between the two edges.
\cite{Son1} Transverse electric field or chemical decoration turn
zigzag graphene NRs to half metal, which makes them good candidates
for spintronics applications. \cite{Son2, APL-kaner, JACS-kaner}
However, metal-free half-metallicity has not been obtained without
the aid of external field or chemical modification.

In this paper we study the geometric and electronic structures of
SiC NRs based on density functional theory (DFT). Spin-polarized
calculations are performed to explore possible magnetic ordering.
The width effects on the properties of SiC NRs are carefully
studied. Half-metallicity is predicted for narrow zigzag SiC NRs.
The stability of SiC NRs are also discussed.

\section{COMPUTATIONAL METHOD AND MODEL SYSTEM}

First principles calculations were performed using the Vienna ab
initio simulation package (VASP). \cite{vasp_1, vasp_2} We described
the interaction between ions and electrons using the projector
augmented wave (PAW) approach. \cite{PAW_1, PAW_2} The Perdew-Wang
functional (PW91) under the generalized gradient approximation (GGA)
was used to describe the exchange correlation interaction.
\cite{PW91_1, PW91_2} During the structure optimizations, all atoms
were fully relaxed until the Hellmann-Feynman forces acting on them
smaller than 0.01 eV/\AA. For nanoribbons, the Brillouin-zone
integrations were performed on a 1$\times$1$\times$11 Monkhorst-Pack
grid. \cite{Monkhorst} 120 uniform $k$-points along the one
dimensional Brillouin zone were used to obtain the band structures
of the SiC NRs. The periodic boundary condition was set with the
vacuum region between two neighboring ribbons larger than 10 \AA.
The phonon band structures were calculated with the general utility
lattice program (GULP) \cite{Gale1, Gale2} using the Tersoff force
field. \cite{Tersoff_1, Tersoff_2}

\section{RESULTS and DISCUSSION}

\begin{figure}
\includegraphics[width=7.5cm]{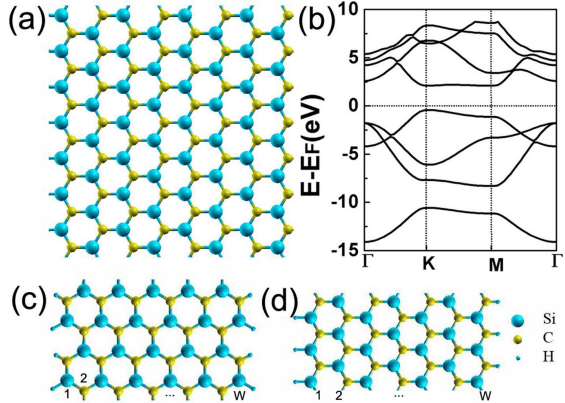}
\caption{(Color online) (a) Structure of a two-dimensional SiC
sheet. (b) Band structure of the graphitic SiC sheet along the
high-symmetry lines of the two-dimensional hexagonal Brillouin zone.
The Fermi level has been set to zero. Structures of an (c) armchair
and a (d) zigzag SiC NRs with a width W. The blue small atoms denote
hydrogen atoms passivate the edge carbon atoms (yellow atoms) and
the edge Silicon atoms (blue big atoms). } \label{fig:geo}
\end{figure}

Before studying the properties of nanoribbons, we first check the
two-dimensional hexagonal SiC sheet. The optimized structure is
shown in Fig. \ref{fig:geo}a, and the obtained Si-C bond length is
1.787 \AA. The electronic band structure shown in Fig.
\ref{fig:geo}b presents a direct energy gap of 2.55 eV at the
\emph{K} point in the hexagonal Brillouin zone, as expected, which
is narrower than the result obtained by self-interaction corrected
DFT calculation. \cite{Baumeier} An accurate first-principles
calculation of band gaps requires a quasiparitcle approach. The
basic physics discovered here however should not be changed.

Similar to graphene NRs \cite{Son1}, we consider two types of SiC
NRs with armchair and zigzag shape edges, as shown in Fig.
\ref{fig:geo}c and \ref{fig:geo}d, respectively.  The nanoribbon
width W is defined as the number of dimer lines along the ribbon
direction for an armchair SiC NR and the number of zigzag chains for
a zigzag SiC NR. All dangling $\sigma$ bonds at the ribbon edges are
saturated by hydrogen atoms. The optimized Si-H and C-H bond lengths
are 1.49 and to 1.09 {\AA}, respectively.

\begin{figure}
\includegraphics[width=7.5cm]{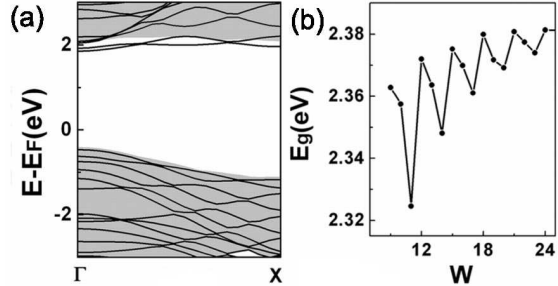}
\caption{(a)Band structure of the armchair SiC NR with W=11. The
projected band structure of a two-dimensional graphene sheet is
shown by shaded areas. (b) The variation of the band gaps of
armchair SiC NRs as a function of the width W.} \label{fig:armchair}
\end{figure}

The band structures for armchair SiC NRs with different widths are
all similar. As an example, the W=11 band structure is plotted in
Fig. \ref{fig:armchair}a. There is a direct band gap at the $\Gamma$
point. As shown in Fig. \ref{fig:armchair}b, the variation of the
energy gap $E_g^W$ as a function of ribbon width W exhibits a
three-family behavior, which is similar to the graphene NRs
\cite{Son1}. The main difference between these two types of armchair
NRs is that the energy gap increases with the ribbon width for SiC
NRs, while it decreases for graphene NRs. Secondly, we have
$E_{g}^{3n}$>$E_{g}^{3n+1}$>$E_{g}^{3n+2}$ for SiC NRs compared to
$E_{g}^{3n+1}$>$E_{g}^{3n}$>$E_{g}^{3n+2}$ for graphene NRs.

\begin{figure}
\includegraphics[width=8.5cm]{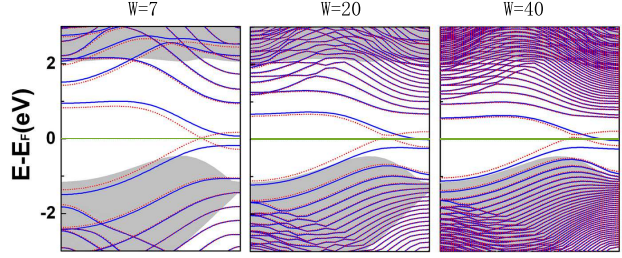}
\caption{(Color online) Spin-polarized band structures of the zigzag
SiC NRs with W=7, 20, and 40. Dotted red lines are spin-up bands,
and solid blue lines are spin-down bands. The projected band
structure of a two-dimensional graphene sheet is shown by shaded
areas.} \label{fig:zigzag}
\end{figure}

For zigzag SiC NRs, we first check the band structure of the W=7 NR,
as shown in Fig. \ref{fig:zigzag}. There are four bands close to the
Fermi level. The two bands in the spin-up channel cross each other,
and both cross the Fermi level. Therefore, the spin-up energy gap is
zero. In the spin-down channel, the two bands close to the Fermi
level form a direct gap at the $X$ point. We name the higher of
these two bands as $H$, the lower as $L$, and the gap between them
as $\Delta^{HL}$. When W=7, the $H$ band is totally unoccupied, and
the $L$ band is totally occupied. The spin-down energy gap is thus
equal to $\Delta^{HL}$, i. e. 0.25 eV. And the zigzag SiC NR with
W=7 presents a half-metallic electronic structure.

\begin{figure}
\includegraphics[width=7.5cm]{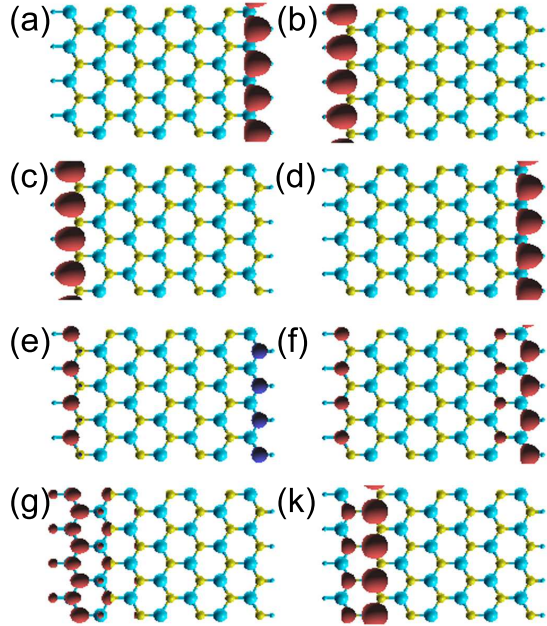}
\hspace{10cm} \caption{(Color online) Densities for the W=7 zigzag
SiC NR. The charge densities of the lowest unoccupied states at $X$
point for the (a) spin-up and (b) spin-down channels. The charge
densities of the highest occupied states at $X$ point for the (c)
spin-up and (d) spin-down channels. (e) The spin density. (f) The
partial charge density with energy range [$E_v$:$E_f$], where $E_v$
is the valence band maximum of the two dimensional SiC sheet and
$E_f$ is the Fermi level of the ribbon. The charge density of the
$H$+1 band at the (g) $\Gamma$ and (h) $X$ points.}
\label{fig:charge}
\end{figure}

The four bands close to the Fermi level are not totally covered by
the projected band structure of the two-dimensional SiC sheet, which
strongly suggests the edge state characters of these four bands. As
shown in Fig. \ref{fig:charge}, at the $X$ point, all the four bands
close to the Fermi level correspond to edge states.

There are local magnetic moments on the edge atoms, and their
orientation are antiparallel between the two edges (Fig.
\ref{fig:charge}e). However, we note that the net magnetic moments
in a unit cell is not zero (refer to table \ref{tbl:gap}).
Therefore, the two edges are not antiferromagnetically coupled as in
zigzag graphene NRs, and they are ferrimagnetically coupled. To
consider the charge transfer between the two edge, we plot the
partial charge density from the valence band maximum of the two
dimensional SiC sheet to the Fermi level in Fig. \ref{fig:charge}f.
We can clearly see a charge transfer from the Si edge to the C edge.

As shown in Table \ref{tbl:gap}, the band gap between the $H$ and
$L$ bands ($\Delta^{HL}$) decreases with the increase of the ribbon
width. We also observed the sink of the $H$ band minimum at the $X$
point with the increase of ribbon widths. When W larger than
$\sim$20, the $H$ band cross with the Fermi level, and the energy
gap in the spin-down channel close. The nanoribbons thus turn to
metal from half metal. In the spin-up channel at the $X$ point, the
occupied state comes from the Si edge, and the unoccupied state
comes from the C edge, which is opposite to the spin-down channel.
The band cross in the spin-up channel leads to a net electron
transfer from the Si edge to the C edge. Therefore, the sink of the
$H$ band at the $X$ point decrease the charge polarization between
the two edges. This is consistent with our previous study \cite{BCN}
on the CBN nanoribbon: the charge polarization decreases when the
ribbon width increases.

\begin{table}[!hbp]
\caption{The energy gap between the $L$ and $H$ bands
($\Delta^{HL}$)and the magnetic moments per cell ($M$) of the widths
($W$) for the zigzag SiC NRs. }\label{tbl:gap}
\begin{tabular}{lp{0.3em}ccccccccc}
\hline \hline
 W       & & $5$ & $6$ & $7$ & $8$ & $9$ & $10$ & $15$  \\
\hline $\Delta^{HL}$(eV) & & 0.39  & 0.34 & 0.25 &  0.24 & 0.24 & 0.24 & 0.085  \\
$M({\mu}_B)$ & & 0.020 & 0.023 & 0.033 & 0.042 & 0.048 & 0.051 & 0.059  \\
\hline \hline
 \end{tabular}
\end{table}

We note that the two dimensional SiC sheet is an insulator with a
big gap. However, for zigzag SiC nanoribbon with W as large as 40,
we still get metallic band structure. For armchair SiC nanoribbons,
as shown in Fig. \ref{fig:armchair} our calculated band gaps also
converge to a value smaller than the bulk gap (2.55 eV). This is due
to the edge states. Similar behavior has been observed for boron
nitride nanoribbons. \cite{BN} There are many bands outside the
projected band structure of two dimensional SiC sheet. In Fig.
\ref{fig:charge}, we plot the density of the $H$+1 band at the
$\Gamma$ and $X$ point. Although they are not edge states, they are
localized on part of the ribbon.

\begin{figure}
\includegraphics[width=7.5cm]{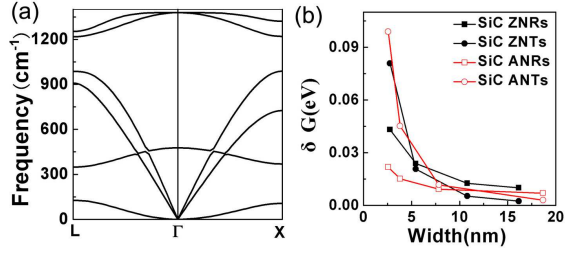}
\hspace{10cm} \caption{(Color online) (a)The phonon band structures
of the 2D SiC graphitic sheet.(b)the Gibbs free energy of formation
of zigzag SiC NRs (SiC ZNRs) and armchair SiC NTs (SiC ANRs) and
corresponding nanotbues (SiC ANRs and SiC ZNTs). The width refers to
that of SiC NRs.} \label{fig:stable}
\end{figure}

To develop applications for SiC NRs, it is important to study their
stability. First, we calculate the phonon band structures of the two
dimensional SiC sheet. As shown in Fig. \ref{fig:stable}, there is
no imaginary frequency exists. The calculated frequency of the
highest optical phonon at $\Gamma$ point is 1378.5 cm$^{-1}$. For
SiC NRs and corresponding nanotubes, we define the Gibbs free energy
of formation as a function of the chemical potential of the silicon
carbide and hydrogen species.
\begin{equation}
\delta{G}=E_{c}-n_H\times \mu_H-n_{SiC}\times \mu_{SiC}
\end{equation}
where $E_{c}$ is the cohesive energy per atom, $n_{SiC(H)}$ and
$\mu_{SiC(H)}$ are the mole fraction and the chemical potential of
silicon carbide (hydrogen), respectively. $\mu_H$ is set to half of
the energy of H$_2$. $\mu_{SiC}$ is the conhensive energy per Si-C
unit of infinite two dimensional SiC sheet. Therefore, the
$\delta{G}$ of 2D SiC sheet is zero by definition. The comparison of
$\delta{G}$ between SiC NRs and NTs with the same numbers of Si and
C atoms are shown in Fig. \ref{fig:stable}. We can see that there is
a $\delta{G}$ crossover between the SiC NRs and the SiC NTs. The
narrower zigzag SiC NRs are even more stable than the corresponding
NTs.

\section{CONCLUSIONS}

In conclusion, We have investigated the electronic and magnetic
properties of SiC NRs with armchair and zigzag shaped edges.  We
show that the armchair SiC NRs always have a wide band gap for all
width. The zigzag SiC NRs with the width smaller than $\sim4$ nm
present half metallic behavior. This makes narrow zigzag SiC
nanoribbon a great candidate for spintronics application, since
neither transverse electric field nor chemical modification is
needed to obtain half-matallicity. The Gibbs free energy of
formation of SiC NRs are similar to that of previously studied SiC
NTs, narrow nanoribbons are even more stable than the experimentally
synthesized SiC nanotubes, which suggests it is practical to
synthesize SiC nanoribbons.

\begin{acknowledgements}

This work was partially supported by the National Natural Science
Foundation of China under Grand Nos. 10674121, 20533030, 10574119,
and 50121202, by National Key Basic Research Program under Grant No.
2006CB922004, by the USTC-HP HPC project, and by the SCCAS and
Shanghai Supercomputer Center.
\end{acknowledgements}

\end{document}